\documentclass{llncs}

\usepackage{llncsdoc}
\usepackage{graphicx}
\usepackage{amsmath,amssymb}
\usepackage{tikz}

\begin{document}

\title{Privacy-Preserving Protocols \\ for Eigenvector Computation}

\author{Manas A. Pathak and Bhiksha Raj}

\institute{
  Carnegie Mellon University \\ 
  \texttt{\{manasp, bhiksha\}@cs.cmu.edu}
}

\maketitle

\begin{abstract}
In this paper, we present a protocol for computing the principal
eigenvector of a collection of data matrices belonging to multiple
semi-honest parties with privacy constraints. Our proposed protocol is
based on secure multi-party computation with a semi-honest arbitrator
who deals with data encrypted by the other parties using an additive
homomorphic cryptosystem. We augment the protocol with randomization
and obfuscation to make it difficult for any party to estimate
properties of the data belonging to other parties from the
intermediate steps. The previous approaches towards this problem were
based on expensive QR decomposition of correlation matrices, we
present an efficient algorithm using the power iteration method. We
analyze the protocol for correctness, security, and efficiency.

\end{abstract}

\section{Introduction}
Eigenvector computation is one of the most basic tools of data
analysis. In any multivariate dataset, the eigenvectors provide
information about key trends in the data, as well as the relative
importance of the different variables. These find use in a diverse set
of applications, including principal component
analysis~\cite{Massy1965Principal},  collaborative
filtering~\cite{Goldberg01} and PageRank~\cite{pagerank}.
Not all eigenvectors of the data are equally important; only those
corresponding to the highest eigenvalues are used as
representations of trends in the data. The most important eigenvector
is the {\em principal} eigenvector corresponding to the maximum
eigenvalue. 

In many scenarios, the entity that actually computes the eigenvectors
is different from the entities that possess the data. For instance, a
data mining agency may desire to compute the eigenvectors of a
distributed set of records, or an enterprise providing recommendations
may want to compute eigenvectors from the personal ratings of
subscribers to facilitate making recommendations to new customers. We
will refer to such entities as \emph{arbitrators}.
Computation of eigenvectors requires the knowledge of either the data
from the individual parties or the correlation matrix derived from it.
The parties that hold the data may however consider them private and
be unwilling to expose any aspect of their individual data to either
the arbitrator or to other parties, while being agreeable, in
principle, to contribute to the computation of a global trend.  As a
result, we require a privacy preserving algorithm that can compute the
eigenvectors of the aggregate data while maintaining the necessary
privacy of the individual data providers.

The common approach to this type of problem is to obfuscate individual
data through controlled randomization~\cite{Evfimievski02}. However,
since we desire our estimates to be exact, simple randomization
methods that merely ensure accuracy in the mean cannot be
employed. Han, \emph{et al.}~\cite{HanNY09} address the problem by
computing the complete QR decomposition~\cite{GVL89} of privately
shared data using cryptographic primitives. This enables all parties
to collaboratively compute the complete set of global eigenvectors but
does not truly hide the data from individual sources. Given the
complete set of eigenvectors and eigenvalues provided by the QR
decomposition, any party can reverse engineer the correlation matrix
for the data from the remaining parties and compute trends among them.
Canny~\cite{Canny02} present a different distributed approach that
does employ an arbitrator, in their case a \emph{blackboard}, however
although individual data instances are hidden, both the arbitrator and
individual parties have access to all aggregated individual stages of
the computation and the final result is public, which is much less
stringent than our privacy constraints.

In this paper, we propose a new privacy-preserving protocol for shared
computation of the principal eigenvector of a distributed collection
of privately held data. The algorithm is designed such that the
individual parties, whom we will refer to as ``Alice'' and ``Bob''
learn nothing about each others' data, and only learn the degree to
which their own data follow the global trend indicated by the
principal eigenvector. The arbitrator, who we call ``Trent'',
coordinates the computation but learns nothing about the data of the
individual parties besides the principal eigenvector which he receives
at the end of the computation.  In our presentation, for simplicity,
we initially consider two parties each having an individual data
matrix. Later we show that the protocol can be naturally generalized
to $N$ parties.  As the $N$ parties communicate only with Trent in a
star network topology with $O(N)$ data transmissions, this is much
more efficient than the $O(N^2)$ data transmission cost if all parties
communicated with each other in a fully connected network.  The data
may be split in two possible ways: along data instances or
features. In this work, we principally consider the data-split
case. However, our algorithm is easily applied to feature
split data as well.

We use the power iteration method~\cite{GVL89} to compute the
principal eigenvector. The arbitrator Trent
introduces a combination of homomorphic encryption~\cite{Paillier99},
randomization, and obfuscation to ensure that
the computation preserves privacy.  The algorithm assumes the parties
to be \emph{semi-honest}. While they are assumed to follow the
protocol correctly and refrain from using falsified data as input,
they may record and analyze the intermediate results obtained while
following the protocol in order to to gain as much information as
possible. It is required that no party colludes with Trent as this
will compromise the privacy of the protocol.

The computational requirements of the algorithm are the same as that
of the power iteration method. In addition, each iteration requires
the encryption and decryption of two $k$ dimensional vectors, where
$k$ is the dimensionality of the data, as well as transmission of the
encrypted vectors to and from Trent. Nevertheless, the encryption and
transmission overhead, which is linear in $k$, may be expected to be
significantly lower than the calculating the QR decomposition or
similar methods which require repeated transmission of entire
matrices. In general, the computational cost of the protocol is
dependent on the degree of security we desire as required by the
application.

\section{Preliminaries}

\subsection{Power Iteration Method}
The power iteration method~\cite{GVL89} is an algorithm to find the
principal eigenvector and its associated eigenvalue for square
matrices. To simplify explanation, we assume that the matrix is
diagonalizable with real eigenvalues, although the algorithm is
applicable to general square matrices as well \cite{Sewell05}.  Let
$A$ be a size $N \times N$ matrix whose eigenvalues are
$\lambda_1,\ldots,\lambda_N$.

The power iteration method computes the principal eigenvector of $A$
through the iteration
\[ x_{n+1} \leftarrow \frac{Ax_n}{\|Ax_n\|}, \]
where $x_n$ is a $N$ dimensional vector. If the principal eigenvalue is
unique, the series $\omega_n = A^nx_0$ is guaranteed to converge to a
scaling of the principal eigenvector.  In the standard algorithm,
$\ell_2$ normalization is used to prevent the magnitude
of the vector from overflow and underflow. Other normalization factors
can also be used if they do not change the limit of the series.

We assume wlog that $|\lambda_1| \ge \cdots \ge |\lambda_N| \ge 0$.
Let $v_i$ be the normalized eigenvector corresponding to $\lambda_i$.
Since $A$ is assumed to be diagonalizable, the eigenvectors
$\{v_1,\ldots,v_N\}$ create a basis for $\mathbb{R}^{N}$. For
unique values of $c_i \in \mathbb{R}^N$, any vector $x_0 \in
\mathbb{R}^N$ can be written as $x_0 = \sum_{i=1}^{N} c_i v_i$.
It can be shown that $\frac{1}{|\lambda_{1}|^{n}}A^{n}x_{0}$ is
asymptotically equal to $c_1v_1$ which forms the basis of the power
iteration method and the convergence rate of the algorithm is
$\left|\frac{\lambda_2}{\lambda_1}\right|$. The algorithm converges
quickly when there is no eigenvalue close in magnitude to the
principal eigenvalue.

\subsection{Homomorphic Encryption}\label{sec:homomorphic}
A homomorphic encryption algorithm allows for operations to be perform
on the encrypted data without requiring to know the unencrypted
values. If $\cdot$ and $+$ are two operators and $x$ and
$y$ are two plaintext elements, a homomorphic encryption function $E$
satisfies
\[ E[x] \cdot E[y] = E[x + y]. \]
In this work, we use the additive homomorphic Paillier asymmetric key 
cryptosystem~\cite{Paillier99}.

\section{Privacy Preserving Protocol}\label{sec:protocol}
\subsection{Data Setup and Privacy Requirements}

We formally define the problem, in which multiple parties, try to
compute the principal eigenvector over their collectively held datasets
without disclosing any information to each other. For simplicity, we
describe the problem with two parties, Alice and Bob; and later show
that the algorithm is easily extended to multiple parties.

The parties Alice and Bob are assumed to be \emph{semi-honest} which
means that the parties will follow the steps of the protocol correctly
and will not try to cheat by passing falsified data aimed at
extracting information about other parties.  The parties are assumed
to be curious in the sense that they may record the outcomes of all
intermediate steps of the protocol to extract any possible
information.  The protocol is coordinated by the semi-honest
arbitrator Trent. Alice and Bob communicate directly with Trent rather
than each other. Trent performs all the intermediate computations and
transfers the results to each party.  Although Trent is trusted not to
collude with other parties, it is important to note that the parties
do not trust Trent with their data and intend to prevent him from
being able to see it. Alice and Bob hide information by using a shared
key cryptosystem to send only encrypted data to Trent.

We assume that both the datasets can be represented as matrices
in which columns and rows correspond to the data samples and the
features, respectively. For instance, the individual email collections
of Alice and Bob are represented as matrices $A$ and $B$ respectively,
in which the columns correspond to the emails, and the rows correspond
to the words. The entries of these matrices represent the frequency of
occurrence of a given word in a given email.  The combined dataset may
be split between Alice and Bob in two possible ways. In a \emph{data}
split, both Alice and Bob have a disjoint set of data samples with the
same features. The aggregate dataset is obtained by concatenating
columns given by the data matrix $M = \begin{bmatrix} A &
  B \end{bmatrix}$ and correlation matrix $M^TM$. In a \emph{feature}
split, Alice and Bob have different features of the same data. The
aggregate data matrix $M$ is obtained by concatenating rows given by
the data matrix $M = \begin{bmatrix} A \\ B \end{bmatrix}$ and
correlation matrix $MM^T$.  If $v$ is an eigenvector of $M^TM$ with a
non-zero eigenvalue $\lambda$, we have
\[M^TM v = \lambda v ~\Rightarrow~ MM^TMv = \lambda M v. \]
Therefore, $Mv \ne 0$ is the eigenvector of $MM^T$ with eigenvalue
$\lambda$. Similarly, any eigenvector of horizontally split data
$MM^T$ associated with a non-zero eigenvalue is an eigenvector of
vertically split data $M^TM$ corresponding to the same eigenvalue.
Hence, we mainly deal with calculating the principal eigenvector of the
vertically split data. In practice the correlation matrix that has
the smaller size should be used to reduce the computational cost of
eigen-decomposition algorithms.

For vertical data split, if Alice's data $A$ is of size $k \times m$
and Bob's data $B$ is of size $k \times n$, the combined data matrix
will be $M_{k \times (m+n)}$. The correlation matrix of size $(m+n)
\times (m+n)$ is given by
\[M^TM = \begin{bmatrix}
  A^TA ~&~ A^TB \\
  B^TA ~&~ B^TB
\end{bmatrix}.\]

\subsection{The Basic Protocol}\label{sec:basic}

The power iteration algorithm computes the principal eigenvector of
$M^TM$ by updating and normalizing the vector $x_t$ until
convergence. Starting with a random vector $x_0$, we calculate 
\[x_{i+1} = \frac{M^TM ~x_i}{\|M^TM ~x_i\|}.\]
For privacy, we split the vector $x_i$ into two parts, $\alpha_i$ and
$\beta_i$. $\alpha_i$ corresponds to the first $m$ components of $x_i$
and $\beta_i$ corresponds to the remaining $n$ components. In each
iteration, we need to securely compute
\begin{align}
  M^TM x_i &= 
  \begin{bmatrix}
    A^TA ~&~ A^TB \\
    B^TA ~&~ B^TB
  \end{bmatrix}
  \begin{bmatrix}
    \alpha_i \\
    \beta_i
  \end{bmatrix} 
  = \begin{bmatrix}
    A^T (A\alpha_i + B\beta_i) \\
    B^T (A\alpha_i + B\beta_i)
  \end{bmatrix}
  = \begin{bmatrix}
      A^T u_i \\
      B^T u_i
    \end{bmatrix} \label{eqn:product}
\end{align}
where $u_i = A\alpha_i+B\beta_i.$ After convergence, $\alpha_i$ and
$\beta_i$ will represent shares held by Alice and Bob of the principal
eigenvector of $M^TM$.

\begin{figure}[ht]
  \centering
  \includegraphics[width=\textwidth]{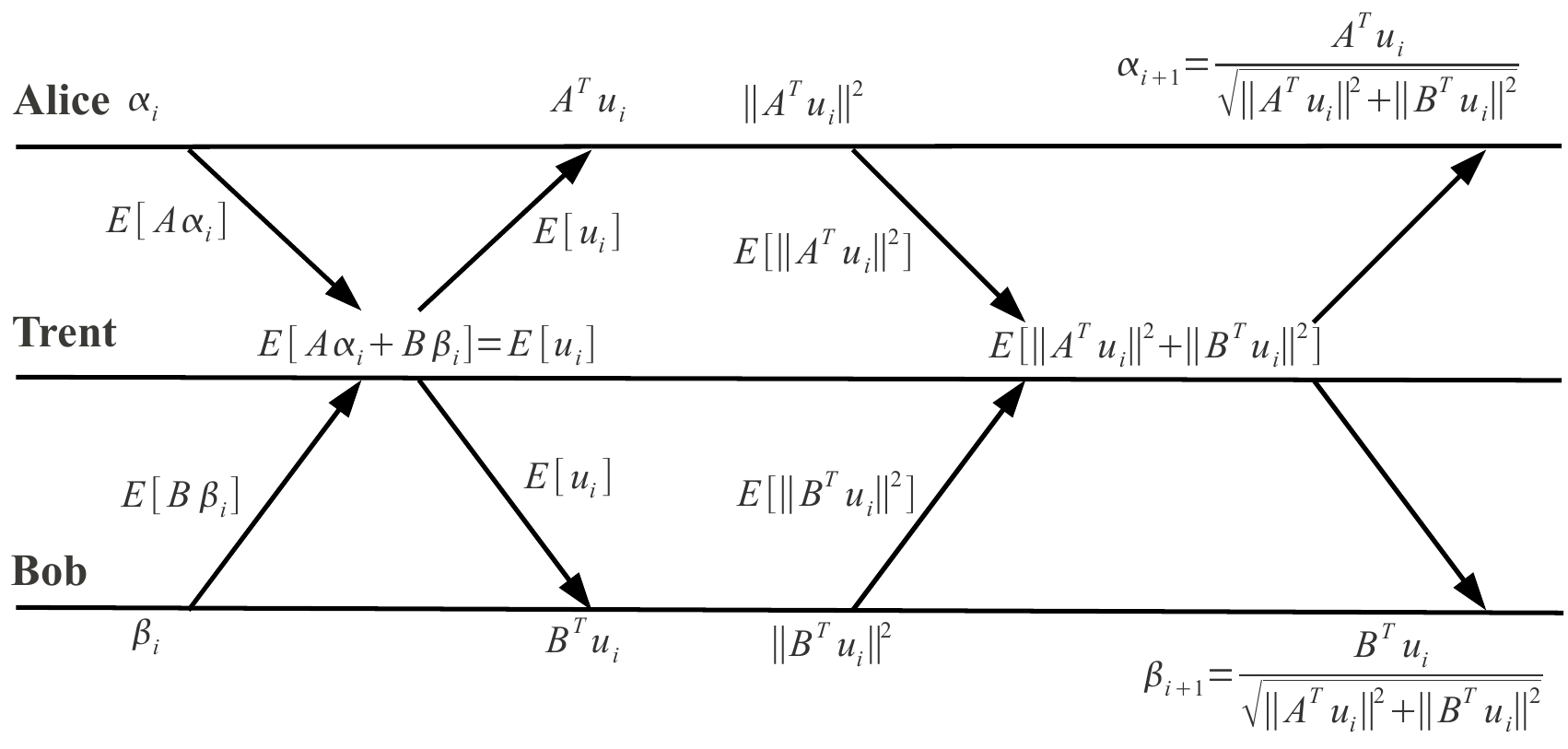}
  \caption{Visual description of the protocol.}
  \label{fig:basic-protocol}
\end{figure}
This now lays the groundwork for us to define a distributed protocol
in which Alice and Bob work only on their portions of the data, while
computing the principal eigenvector of the combined data in
collaboration with a third party Trent.  An iteration of the algorithm
proceeds as illustrated in Fig.~\ref{fig:basic-protocol}. At the
outset Alice and Bob randomly generate component vectors $\alpha_0$
and $\beta_0$ respectively.  At the beginning of the $i^{\rm th}$
iteration, Alice and Bob possess component vectors $\alpha_i$ and
$\beta_i$ respectively. They compute the product of their data and
their corresponding component vectors as $A\alpha_i$ and
$B\beta_i$. To compute $u_i$, Alice and Bob individually transfer
these products to Trent.  Trent adds the contributions from Alice and
Bob by computing
\[u_i = A\alpha_i + B\beta_i.\]
He then transfers $u_i$ back to Alice and Bob,
who then individually compute $A^Tu_i$ and $B^Tu_i$, without
requiring data from one other.
For normalization, Alice and Bob also need to securely compute the term
\begin{align}
  \|M^TM~x_i\| = \sqrt{\|A^Tu_i\|^2 + \|B^Tu_i\|^2}.
\end{align}
Again, Alice and Bob compute the individual terms $\|A^Tu_i\|^2$ and
$\|B^Tu_i\|^2$ respectively and transfer it to Trent. As earlier, Trent
computes the sum
\[ \|A^Tu_i\|^2 + \|B^Tu_i\|^2 \]
and transfers it back to Alice and Bob. Finally, Alice and Bob
respectively update $\alpha$ and $\beta$ vectors as
\begin{align}
  &u_i = A\alpha_i + B\beta_i, \nonumber \\
  &\alpha_{i+1} = \frac{A^T u_i}{\sqrt{\|A^Tu_i\|^2 + \|B^Tu_i\|^2}}, \nonumber \\
  &\beta_{i+1} = \frac{B^T u_i}{\sqrt{\|A^Tu_i\|^2 + \|B^Tu_i\|^2}}.
\end{align}
The algorithm terminates when the $\alpha$ and $\beta$ vectors converge.

\subsection{Making the Protocol More Secure}\label{sec:securing}
The basic protocol described above is provably correct. After
convergence, Alice and Bob end up with the principal eigenvector of
the row space of the combined data, as well as concatenative shares of
the column space which Trent can gather to compute the principal
eigenvector. However the protocol is not secure; Alice and Bob obtain
sufficient information about properties of each others' data matrices,
such as their column spaces, null spaces, and correlation matrices. We
present a series of modifications to the basic protocol so that such
information is not revealed.

\subsubsection{Homomorphic Encryption: Securing the data from Trent.}

The central objective of the protocol is to prevent Trent from learning anything
about either the individual data sets or the combined data other than the
principal eigenvector of the combined data. Trent receives a
series of partial results of the form $AA^T u$, $BB^T u$ and $MM^T u$. By
analyzing these results, he can potentially determine the entire column
spaces of Alice and Bob as well as the combined data. To prevent this, we
employ an additive homomorphic cryptosystem introduced in Section
\ref{sec:homomorphic}.

At the beginning of the protocol, Alice and Bob obtain a shared public
key/private key pair for an additive homomorphic cryptosystem from an
authenticating authority. The public key is also known to Trent who,
however, does not know the private key; While he can encrypt data, he
cannot decrypt it. Alice and Bob encrypt all transmissions to Trent, at the first
transmission step of each iteration Trent receives the encrypted
inputs $E[A\alpha_i]$ and $E[B\beta_i]$. He multiplies the two element
by element to compute $E[A\alpha_i]\cdot E[B\beta_i] = E[A\alpha_i +
  B\beta_i] = E[u_i]$.  He returns $E[u_i]$ to both Alice and Bob who
decrypt it with their private key to obtain $u_i$. In the second
transmission step of each iteration, Alice and Bob send
$E[\|A^Tu_i\|^2]$ and $E[\|B^Tu_i\|^2]$ respectively to Trent, who
computes the encrypted sum
\[ E\left[\|A^Tu_i\|^2\right] \cdot E\left[\|B^Tu_i\|^2\right]
= E\left[\|A^Tu_i\|^2 + \|B^Tu_i\|^2\right] \]
and transfers it back to Alice and Bob, who then decrypt it to obtain
$\|A^Tu_i\|^2 + \|B^Tu_i\|^2$, which is required for normalization.

This modification does not change the actual computation of the power
iterations in any manner. Thus the procedure remains as correct as
before, except that Trent now no longer has any access to any of the
intermediate computations. At the termination of the algorithm he can
now receive the converged values of $\alpha$ and $\beta$ from Alice
and Bob, who will send it in clear text.

\subsubsection{Random Scaling: Securing the Column Spaces.}\label{sec:randomization}

After Alice and Bob receive $u_i = A\alpha_i + B\beta_i$ from Trent,
Alice can calculate $u_i - A\alpha_i = B\beta_i$ and Bob can calculate
$u_i - B\beta_i = A\alpha_i$.  After a sufficient number of
iterations, particularly in the early stages of the computation (when
$u_i$ has not yet converged) Alice can find the column space of $B$
and Bob can find the column space of $A$. Similarly, by subtracting
their share from the normalization term returned by Trent, Alice and
Bob are able to find $\|B^Tu_i\|^2$ and $\|A^Tu_i\|^2$ respectively.

In order to prevent this, Trent multiplies $u_i$ with a randomly generated
scaling term $r_i$ that he does not share with anyone. Trent computes
\[ \left(E[A\alpha_i] \cdot E[B\beta_i]\right)^{r_i} = E[r_i (A\alpha_i + B\beta_i)] = E[r_iu_i] \]
by performing element-wise exponentiation of the encrypted vector by
$r_i$ and transfers $r_iu_i$ to Alice and Bob. By using a different
value of $r_i$ at each iteration, Trent ensures that Alice and Bob are
not able to calculate $B\beta_i$ and $A\alpha_i$ respectively.
In the second step, Trent scales the normalization constant by $r_i^2$,
\begin{align*}
  \left(E\left[\|A^Tu_i\|^2\right] \cdot E\left[\|B^Tu_i\|^2\right]\right)^{r_i^2} 
  = E\left[r_i^2\left(\|A_i^Tu\|^2 + \|B_i^Tu\|^2\right)\right].
\end{align*}
Normalization causes the $r_i$ factor to cancel out and the update
rules remain unchanged.
\begin{align}
  &u_i = A\alpha_i + B\beta_i, \nonumber \\
  &\alpha_{i+1} = \frac{r_i A^T u_i}{\sqrt{r_i^2\left(\|A^Tu_i\|^2 + \|B^Tu_i\|^2\right)}} = \frac{A^T u_i}{\sqrt{\|A^Tu_i\|^2 + \|B^Tu_i\|^2}}, \nonumber \\
  &\beta_{i+1} = \frac{r_i B^T u_i}{\sqrt{r_i^2\left(\|A^Tu_i\|^2 + \|B^Tu_i\|^2\right)}} = \frac{B^T u_i}{\sqrt{\|A^Tu_i\|^2 + \|B^Tu_i\|^2}}.
\end{align}
The random scaling does not affect the final outcome of the
computation, and the algorithm remains correct as before.

\subsubsection{Data Padding: Securing null spaces.}\label{sec:padding}

In each iteration, Alice observes one vector $r_iu_i = r_i (A\alpha_i
+ B\beta_i)$ in the column space of $M = [A  ~ B]$. Alice can
calculate the \emph{null space} $H(A)$ of $A$, given by 
\[ H(A) = \{x \in \mathbb{R}^m | Ax = 0\} \]
and pre-multiply a non-zero vector $x \in H(A)$ with $r_i u_i$ to calculate
\[x r_i u_i = r_i x (A\alpha_i + B\beta_i) = r_i x B\beta_i. \]
This is a projection of $B\beta_i$, a vector in the column space of
$B$ into the null space $H(A)$. Similarly, Bob can find projections of
$A\alpha_i$ in the null space $H(B)$. While considering the projected
vectors separately will not give away much information, after several
iterations Alice will have a projection of the column space of $B$ on
the null space of $A$, thereby learning about the component's of Bob's
data that lie in her null space.  Bob can similarly learn about the
component's of Alice's data that lie in his null space.

In order to prevent this, Alice participates in the protocol with a
{\em padded} matrix $\begin{bmatrix}A & P_a\end{bmatrix}$ as input
created by concatenating her data matrix $A$ with a random matrix
$P_a = r_a I_{k \times k}$, where $r_a$ is a positive scalar chosen
by Alice. Similarly, Bob uses a padded matrix $\begin{bmatrix}B &
P_b\end{bmatrix}$ created by concatenating his data matrix $B$
with $P_b = r_b I_{k \times k}$, where $r_b$ is a different
positive scalar chosen by Bob. This has the effect of hiding the
null spaces in both their data sets. The following lemma shows
that the eigenvectors of the combined data do not change
when using padded matrices. Please refer to appendix for the proof.

\begin{lemma} \label{lem:padding}
Let $\bar{M}=\begin{bmatrix}M & P\end{bmatrix}$
where $M$ is a $s\times t$ matrix, and $P$
is a $s\times s$ orthogonal matrix. If $\bar{v}=\begin{bmatrix}
v_{t\times1} \\ v'_{s\times1}\end{bmatrix}$
is an eigenvector of $\bar{M}^T\bar{M}$
corresponding to an eigenvalue $\lambda$, then
$v$ is an eigenvector of $M^TM$.
\end{lemma}
While the random factors $r_a$ and $r_b$ prevent Alice and Bob from
estimating the eigenvalues of the data, the computation of principal
eigenvector remains correct as before.

\subsubsection{Obfuscation: Securing Krylov spaces.}\label{sec:modificationot}
 
For a constant $c$, we can show that the vector $u_i = A\alpha_i +
B\beta_i$ is equal to $cMM^T u_{i-1}$. The sequence of vectors $U =
\{u_1$, $u_2$, $u_3$, $\ldots\}$ form the Krylov subspace $(MM^T)^n
u_1$ of the matrix $MM^T$.  Knowledge of this series of vectors can
reveal all eigenvectors of $MM^T$.  Consider $u_0 = c_1 v_1 + c_2 v_2
+ \cdots$, where $v_i$ is the $i^{\rm th}$ eigenvector. If $\lambda_j$
is the $j^{\rm th}$ eigenvalue, we have $u_i = c_1 \lambda_1 v_1 +
c_2 \lambda_2 v_2 + \cdots$.  We assume wlog that the eigenvalues
$\lambda$ are in a descending order, \emph{i.e.}, $\lambda_j \ge
\lambda_k$ for $j < k$.  Let $u_{conv}$ be the normalized converged
value of $u_i$ which is equal to the normalized principal eigenvector
$v_1$.

Let $w_i = u_i - (u_i\cdot u_{conv})u_i$ which can be shown to be
equal to $c_2 \lambda_2 v_2 + c_3 \lambda_3 v_3 + \cdots$,
\emph{i.e.}, a vector with no component along $v_1$. If we perform
power iterations with initial vector $w_1$, the converged vector
$w_{conv}$ will be equal to the eigenvector corresponding to the
second largest eigenvalue.  Hence, once Alice has the converged value,
$u_{conv}$, she can subtract it out of all the stored $u_i$ values and
determine the second principal eigenvector of $MM^T$. She can repeat
the process iteratively to obtain all eigenvectors 
of $MM^T$, although in practice the estimates become noisy very
quickly. As we will show in Section \ref{sec:anal}, the following
modification prevents Alice and Bob from identifying the Krylov space
with any certainty and they are thereby unable to compute the
additional eigenvectors of the combined data.

We introduce a form of obfuscation; we assume
that Trent stores the encrypted results of intermediate steps at every
iteration. After computing $E[r_i u_i]$, Trent either sends this
quantity to Alice and Bob with a probability $p$ or sends a random
vector $E[u_i']$ of the same size ($k \times 1$) with probability
$1-p$. As the encryption key of the cryptosystem is publicly known,
Trent can encrypt the vector $u_i'$. Alice and Bob do not know whether
they are receiving $r_i u_i$ or $u_i'$.  If a random vector is sent,
Trent continues with the protocol, but ignores the terms Alice and Bob
return in the next iteration, $E[A\alpha_{i+1}]$ and
$E[B\beta_{i+1}]$. Instead, he sends the result of a the last
non-random iteration $j$, $E[r_j u_j]$, thereby restarting that
iteration.

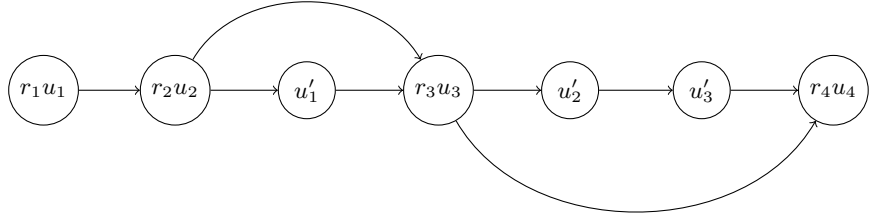
\begin{figure}[ht]
  \centering
  \begin{tikzpicture}
    \node (A1) at (0,0) [circle,draw] {$r_1u_1$};
    \node (A2) at (1.75,0) [circle,draw] {$r_2u_2$};
    \node (A3) at (3.5,0) [circle,draw] {$u_1'$};
    \node (A4) at (5.25,0) [circle,draw] {$r_3u_3$};  
    \node (A5) at (7,0) [circle,draw] {$u_2'$};
    \node (A6) at (8.75,0) [circle,draw] {$u_3'$};
    \node (A7) at (10.5,0) [circle,draw] {$r_4u_4$};
    \draw [->] (A1) to (A2);
    \draw [->] (A2) to (A3);
    \draw [->] (A3) to (A4);
    \draw [->] (A4) to (A5);
    \draw [->] (A5) to (A6);
    \draw [->] (A6) to (A7);
    \draw [->] (A2) to [out=60,in=120] (A4);
    \draw [->] (A4) to [out=-60,in=-120] (A7);
  \end{tikzpicture}
  \caption{An example of the protocol execution with obfuscation.}
  \label{fig:bernoulli}
\end{figure}

This sequence of data sent by Trent is an example of a
Bernoulli Process~\cite{Papoulis84}.
An illustrative example of the protocol is shown in Fig.
\ref{fig:bernoulli}. In the first two iterations, Trent sends valid
vectors $r_1u_1$ and $r_2u_2$ back to Alice and Bob. In the beginning
of the third iteration, Trent receives and computes $E[r_3u_3]$ but
sends a random vector $u_3'$. He ignores what Alice and Bob
send him in the fourth iteration and sends back $E[r_3u_3]$
instead. Trent then stores the vector $E[r_4u_4]$ sent by Alice and
Bob in the fifth iteration and sends a random vector
$u_2'$. Similarly, he ignores the computed vector of the sixth
iteration and sends $u_3'$. Finally, he ignores the computed vector of
the seventh iteration and sends $E[r_4u_4]$.

This modification has two effects -- firstly it prevents Alice and Bob from
identifying the Krylov space with certainty. As a result, they are
now unable to obtain additional eigenvectors from the data.  Secondly,
the protocol essentially obfuscates the projection of the column space
of $B$ on to the null space of $A$ for Alice, and analogously for Bob
by introducing random vectors. As Alice and Bob do not know which vectors
are random, they cannot completely calculate the true projection of
each others data on the null spaces. This is rendered less important
if Alice and Bob pad their data as suggested in the previous
subsection.

Alice and Bob can store the vectors they receive from Trent in each
iteration. By analyzing the distribution of the normalized vectors,
Alice and Bob can identify the random vectors using a simple outlier
detection technique. To prevent this, one possible solution is for
Trent to pick a previously computed value of $r_ju_j$ and add zero
mean noise $e_i$, for instance, sampled from the Gaussian
distribution.
\[ u_i' = r_j u_j + e_i, ~~ e_i \sim \mathcal{N}(0,\sigma^2). \]
Instead of transmitting a perturbation of a previous vector, 
Trent can also use perturbed mean of a few previous $r_j u_j$
with noise. Doing this will create a random vector with the same
distributional properties as the real vectors. The noise variance
parameter $\sigma$ controls the error in identifying the random vector
from the valid vectors and how much error do we want to introduce in
the projected column space.

obfuscation has the effect of increasing the total
computation as every iteration in which Trent sends a random vector is
wasted. In any secure multi-party computation, there is an inherent
trade-off between computation time and the degree of security. The
parameter $p$ which is the probability of Trent sending a non-random
vector allows us to control this at a fine level based on the
application requirements.  As before, introducing obfuscation
does not affect the correctness of the computation -- it does not
modify the values of the non-random vectors $u_i$.

\subsection{Extension to Multiple Parties}\label{sec:multiparty}

As we mentioned before, the protocol can be naturally extended to
multiple parties. Let us consider the case of $N$ parties:
$P_1,\ldots,P_N$ each having data $A_1,\ldots,A_N$ of sizes $k
\times n_1,\ldots,k \times n_N$ respectively. The parties
are interested in computing the principal eigenvector of the combined
data without disclosing anything about their data. We make the same
assumption about the parties and the arbitrator Trent being
\emph{semi-honest}. All the parties except Trent share the decryption
key to the additive homomorphic encryption scheme and the encryption
key is public.

In case of a data split, for the combined data matrix $M =
\begin{bmatrix}A_1 & A_2  & \cdots & A_N\end{bmatrix}$, the
correlation matrix is
\[M^TM =
\begin{bmatrix}
  A_1^TA_1 &  \cdots & A_1^TA_N \\
  \vdots    & \ddots & \vdots  \\
  A_N^TA_1 & \cdots & A_N^TA_N \\
\end{bmatrix}.\]
We split the eigenvector into $N$ parts, $\alpha_1,\ldots,\alpha_N$ of
size $n_1,\ldots,n_N$ respectively, each corresponding to one party.
For simplicity, we describe the basic protocol with homomorphic
encryption; randomization and obfuscation can be easily added
by making the same modifications as we saw in Sections 3.3. One
iteration of the protocol starts with the $i^{\rm th}$ party
computing $A_i\alpha_i$ and transferring to Trent the 
encrypted vector $E[A_i\alpha_i]$. Trent receives this from each party
and computes
\[ \prod_i E\left[A_i\alpha_i\right] = E\left[\sum_i
  A_i\alpha_i\right] = E[u] \]
where $u = \sum_i A_i\alpha_i$, and product
is an element-wise operation. Trent sends
the encrypted vector $E[u]$ back to $P_1,\ldots,P_N$ who decrypt
it and individually compute $A_i^Tu$. The parties individually compute
$\|A_i^Tu\|^2$ and send its encrypted value to Trent. Trent receives
$N$ encrypted scalars $E\left[\|A_i^Tu\|^2\right]$ and calculates the
normalization term
\[ \prod_i E\left[\|A_i^Tu\|^2\right] = E\left[\sum_i \|A_i^Tu\|^2\right] \]
and sends it back to the parties. At the end of the iteration, the party $P_i$
updates $\alpha_i$ as
\begin{align}
  &u = \sum_i A_i\alpha_i^{(old)}, \nonumber \\
  &\alpha_{i}^{(new)} = \frac{A_i^T u}{\sqrt{\sum_i \|A_i^T u\|^2}}.
\end{align}
The algorithm terminates when any one party $P_i$ converges on
$\alpha_i$.

\section{Analysis}\label{sec:anal}
\subsection{Correctness}

The protocol outlined in Section \ref{sec:basic} is provably
correct. The steps introduced in Section \ref{sec:securing} do not
modify the operation and hence the accuracy of the protocol in any
manner.
 
\subsection{Security}

As a consequence of the procedures introduced in Section
\ref{sec:randomization} the row spaces and null spaces of the parties
are hidden from each another. In the multiparty scenario, the protocol
is also robust to collusion between parties with data, although not to
collusion between Trent and any of the other parties. If two parties
out of $N$ collude, they will find information about each other, but
will not learn anything about the data of the remaining $N-2$ parties.

What remains is the information which can be obtained from
the sequence of $u_i$ vectors. Alice receives the following two sets
of matrices:
\[ U = \{u_1, u_2, u_3, \ldots\},~~ U' = \{u'_1, u'_2,\ldots\} \]
representing the outcomes of valid iterations and the random vectors
respectively. In the absence of the random data $U'$, Alice only receives
$U$. As mentioned in Section \ref{sec:modificationot}, $u_i = (MM^T)^i
u_0$ which is a sequence of vectors from the Krylov space
of the matrix $AA^T + BB^T$ sufficient to determine 
all eigenvectors of $MM^T$. For $k$-dimensional data, it
is sufficient to have any sequence of $k$ vectors in $U$ to determine
$MM^T$. Hence, if the vectors in $U$ were not interspersed with the
vectors in $U'$, the algorithm essentially reveals information about
all eigenvectors to all parties.  Furthermore, given a sequence $u_i,
u_{i+1}, u_{i+2},\ldots,u_{i+k-1}$ vectors from $U$, Alice can {\em
  verify} that they are indeed from the Krylov
space.\footnote{if the spectral radius of $MM^T$ is
  1.} Introducing random scaling $r_i u_i$ makes it harder still to verify
Krylov space. While solving for $k$ vectors, Alice and Bob need to
solve for another $k$ parameters $r_1,\ldots,r_k$.

Security is obtained from the following observation: although Alice
can {\em verify} that a given set of vectors forms a sequence in the
Krylov space, she cannot {\em select} them from a larger set without
exhaustive evaluation of all $k$ sets of vectors. If the shortest
sequence of $k$ vectors from the Krylov space is embedded in a longer
sequence of $N$ vectors, Alice needs $N \choose k$ checks to find the
Krylov space, which is a combinatorial problem.

\subsection{Efficiency}
First we analyze the computational time complexity of the protocol. As
the total the number of iterations is data dependent and proportional
to $\left|\frac{\lambda_1}{\lambda_2}\right|$, we analyze the cost per
iteration. The computation is performed by the individual parties in
parallel, though synchronized and the parties also spend time waiting
for intermediate results from other parties. Obfuscation
introduces extra iterations with random data, on average
the number of iterations needed for convergence increase by a
factor of $\frac{1}{p}$, where $p$ is the probability of Trent sending
a non-random vector. As the same operations are performed in an 
iteration with a random vector, its the time complexity would be the same as an
iteration with a non-random vector.

In the $i^{\rm th}$ iteration, Alice and Bob individually need to perform two
matrix multiplications: $A\alpha_i$ and $A^T(A\alpha_i + B\beta_i)$,
$B\beta_i$ and $B^T(A\alpha_i + B\beta_i)$ respectively. The first
part involves multiplication of a $k \times m$ matrix by a $m$
dimensional vector which is $O(km)$ operations for Alice and $O(kn)$
for Bob. The second part involves multiplication of a $m \times k$
matrix by a $k$ dimensional vector which is $O(km)$ operations for
Alice and $O(kn)$ for Bob. Calculating $\|A^T(A\alpha_i +
B\beta_i)\|^2$ involves $O(m)$ operations for Alice and analogously
$O(n)$ operations for Bob. The final step involves only a
normalization by a scalar and can be again done in linear
time, $O(m)$ for Alice and $O(n)$ for Bob. Therefore, total time
complexity of computations performed by Alice and Bob is $O(km) + O(m)
= O(km)$ and $O(kn) + O(n) = O(kn)$ operations respectively.
Trent computes an element-wise product of two $k$ dimensional vectors
$A\alpha_i$ and $B\beta_i$ which is $O(k)$ operations. The
multiplication of two encrypted scalar requires only one
operation, making Trent's total time complexity $O(k)$.

In each iteration, Alice and Bob encrypt and decrypt two vectors and
two scalar normalization terms which is equivalent to performing $k+1$
encryptions and $k+1$ decryptions individually, which is $O(k)$
encryptions and decryptions.

In the $i^{th}$ iteration, Alice and Bob each need to transmit $k$ dimensional
vectors to Trent who computes $E(A\alpha_i + B\beta_i)$ and transmits
it back: involving the transfer of $4k$ elements. Similarly, Alice and
Bob each transmit one scalar norm value to Trent who sends back
another scalar value involving in all the transfer of 4 elements. In
total, each iteration requires the transmission of $4k+4 = O(k)$ data
elements.

To summarize, the time complexity of the protocol per iteration is
$O(km)$ or $O(kn)$ operations whichever is larger, $O(k)$ encryptions
and decryptions, and $O(k)$ transmissions. In practice, each
individual encryption/decryption and data transmission take much
longer than performing computation operation.

\section{Conclusion}
In this paper, we proposed a protocol for computing the principal
eigenvector of the combined data shared by multiple parties
coordinated by a semi-honest arbitrator Trent. The data matrices
belonging to individual parties and
correlation matrix of the combined data is protected and cannot be
reconstructed. We used randomization, data padding, and obfuscation
to hide the information which the parties can learn from the
intermediate results. The computational cost for each party is
$O(km)$ where $k$ is the number of features and $m$ data
instances along with $O(k)$ encryption and decryption operations and
$O(k)$ data transfer operations.

Potential future work include extending the protocol to finding the
complete singular value decomposition, particularly with efficient
algorithms like thin SVD~\cite{Brand2006}. Some of the techniques such
as data padding and obfuscation can be applied to other problems as
well. We are working towards a unified theoretical model for applying
and analyzing these techniques in general.

\bibliographystyle{abbrv}
\bibliography{ppsvd}

\section{Appendix}
\begin{proof}[Lemma \ref{lem:padding}]
We have,
\[ \bar{M}^T\bar{M} = \begin{bmatrix}
    M^TM ~&~ M^TP \\
    P^TM ~&~ I \end{bmatrix}. \]
Multiplying by the eigenvector $\bar{v}=\begin{bmatrix}
v_{t\times1} \\ v'_{s\times1}\end{bmatrix}$ gives us
\[  \bar{M}^T\bar{M} \begin{bmatrix}v \\ v'\end{bmatrix}
    = \begin{bmatrix}
      M^TMv + M^TPv' \\
      P^TMv + v' \end{bmatrix}
    = \lambda \begin{bmatrix}v \\ v'\end{bmatrix}. \]
Therefore, 
\begin{align}
  M^TMv + M^TPv' &= \lambda v, \label{eq:Lem1} \\
  P^TMv + v' &= \lambda v'. \label{eq:Lem2}
\end{align}

Since $\lambda\neq 1$, Equation \eqref{eq:Lem2} implies $v' =
\frac{1}{\lambda - 1}P^TMv$. Substituting this into Equation
\eqref{eq:Lem1} and the orthogonality of $P$ gives us
\[ M^TMv + \frac{1}{\lambda - 1}M^TPP^TMv 
= \frac{\lambda}{\lambda - 1}M^TMv = \lambda v. \]
Hence, $M^TMv=(\lambda-1)v$. \\ 
\qed
\end{proof}

\end{document}